%% file: main.tex
\documentclass[sigconf,screen]{acmart}

\AtBeginDocument{%
  }

\copyrightyear{2026}
\acmYear{2026}
\setcopyright{cc}
\setcctype{by-nc-nd}
\acmConference[ASE '26]{Proceedings of the 41st IEEE/ACM International Conference on Automated Software Engineering}{October 12--16, 2026}{Munich, Germany}
\acmBooktitle{Proceedings of the 41st IEEE/ACM International Conference on Automated Software Engineering (ASE '26), October 12--16, 2026, Munich, Germany}
\acmDOI{10.1145/3832783.3834468}
\acmISBN{979-8-4007-2882-2/2026/10}

\usepackage{tcolorbox}
\usepackage{xspace}
\usepackage{booktabs}
\usepackage{pbalance}
\usepackage{graphicx}
\usepackage{subcaption}
\usepackage{caption}  
\usepackage{placeins}

\usepackage[export]{adjustbox}

\usepackage{tabularx}
\usepackage{colortbl}
\usepackage{multirow}
\usepackage{arydshln}
\usepackage{booktabs}
\usepackage[table]{xcolor}
\usepackage{array}

\def\eg{\textit{e.g.},~}

\def\namerm{KRCA}
\def\name{\textit{\namerm}\xspace}

\begin{document}

\title{KRCA: An Efficient Root Cause Analysis System in \\ Hyper-scale Microservice Systems via Agentic AI}

% ========================================================= 
% Nankai University 
% ========================================================= 
\author{Jiamin Jiang} \authornote{Work done during an internship at Kuaishou Technology.} 
\affiliation{\institution{Nankai University} \city{Tianjin} \country{China} } 
\email{jiangjiamin@mail.nankai.edu.cn} 

\author{Jingfei Feng} 
\affiliation{\institution{Nankai University} \city{Tianjin} \country{China} } 
\email{fengjingfei@mail.nankai.edu.cn} 

\author{Yu Luo} 
\affiliation{
\institution{Nankai University} \city{Tianjin} \country{China} } 
\email{luoyu@mail.nankai.edu.cn} 

\author{Qingliang Zhang} 
\affiliation{\institution{Nankai University} \city{Tianjin} \country{China} } \email{zhangqingliang@mail.nankai.edu.cn} 

\author{Yongqian Sun}  \authornote{Corresponding author.} 
\affiliation{\institution{Nankai University} \city{Tianjin} \country{China} } 
\email{sunyongqian@nankai.edu.cn} 

\author{Wenwei Gu} 
\affiliation{\institution{Nankai University} \city{Tianjin} \country{China} } 
\email{wwgu@nankai.edu.cn} 

\author{Shenglin Zhang} 
\affiliation{\institution{Nankai University} \city{Tianjin} \country{China} } 
\email{zhangsl@nankai.edu.cn} 

% ========================================================= 
% Kuaishou Technology 
% ========================================================= 

\author{Tianyu Cui} 
\affiliation{\institution{Kuaishou Technology} \city{Beijing} \country{China} } 
\email{cuitianyu@kuaishou.com} 

\author{Yao Wu} 
\affiliation{\institution{Kuaishou Technology} \city{Beijing} \country{China} } 
\email{wuyao05@kuaishou.com} 

\author{Jielong Huang} 
\affiliation{\institution{Kuaishou Technology} \city{Beijing} \country{China} } 

\email{zhangsong08@kuaishou.com} 

\author{Nan Qi} \affiliation{\institution{Kuaishou Technology} \city{Beijing} \country{China} } \email{qinan03@kuaishou.com} 

% ========================================================= 
% Tsinghua University 
% ========================================================= 
\author{Dan Pei} 
\affiliation{\institution{Tsinghua University} \city{Beijing} \country{China} } 
\email{peidan@tsinghua.edu.cn}

\renewcommand{\shortauthors}{Jiang et al.}

\begin{abstract}
  Hyper-scale microservice systems have become the standard infrastructure for large-scale Internet companies. These systems consist of numerous loosely coupled microservices that evolve independently through continuous development and deployment. Such complexity makes failures unavoidable, necessitating efficient Root Cause Analysis (RCA) to help Site Reliability Engineers (SREs) quickly localize root cause services and classify failure types. However, existing RCA methods often struggle to adapt to the extreme dynamism and massive scale of these systems. In this paper, we present \name, an end-to-end RCA system designed for hyper-scale microservice systems. To manage the vast search space, \name employs a multi-stage pipeline that begins with an API-level drilldown to isolate suspicious services. It then instantiates a skeleton-based causal graph from anomalous metrics to serve as a high-recall structural prior, before utilizing a memory-augmented multi-agent framework to verify causality and generate the final failure report. By combining structured causal constraints with multi-agent reasoning, \name balances diagnostic accuracy with the efficiency requirements of real-time production use. Experimental results show that \name achieves AC@1 scores of 0.88 and 0.79 for root cause service localization and failure type classification, outperforming the strongest baseline by at least 31\% in absolute gains. \name has been deployed in Kuaishou's production environment for over six months, reducing the average diagnosis time by 77.3\%.
\end{abstract}




\maketitle

\section{Introduction}\label{intro}
\input{Chapter/Intro}

\section{BackGround and Motivation}

\input{Chapter/Motivation}

\section{System Design}
\input{Chapter/SystemDesign}

\section{Evaluation}
\input{Chapter/Evaluation}

\section{Discussion}
\input{Chapter/Discussion}

\section{Related Work}
\input{Chapter/Related_work}

\section{Conclusion}
\input{Chapter/Conclusion}

\section*{Data Availability Statement}
The datasets used in this study are not publicly available because they contain highly sensitive company-wide production data and proprietary infrastructure information. Due to confidentiality, security, and compliance requirements, these materials cannot be released. The source code is currently undergoing the company’s internal open-source review process and will be released at \url{https://anonymous.4open.science/r/KRCA-0FCC} after desensitization and approval.

\section*{Acknowledgments}
We sincerely thank the anonymous shepherd and reviewers for their comments and feedback. This work is supported by the National Natural Science Foundation of China (62272249, 62302244), the Fundamental Research Funds for the Central Universities (XXX-63253249), and the Tianjin Key Research and Development Program (Grant No. 25YFYFFG00690).

% \clearpage
% \printbibliography
\bibliographystyle{ACM-Reference-Format}
\bibliography{ref}
\end{document}

%% file: Chapter/Intro.tex
\begin{figure}[t]
    \centering
    \includegraphics[width=\linewidth]{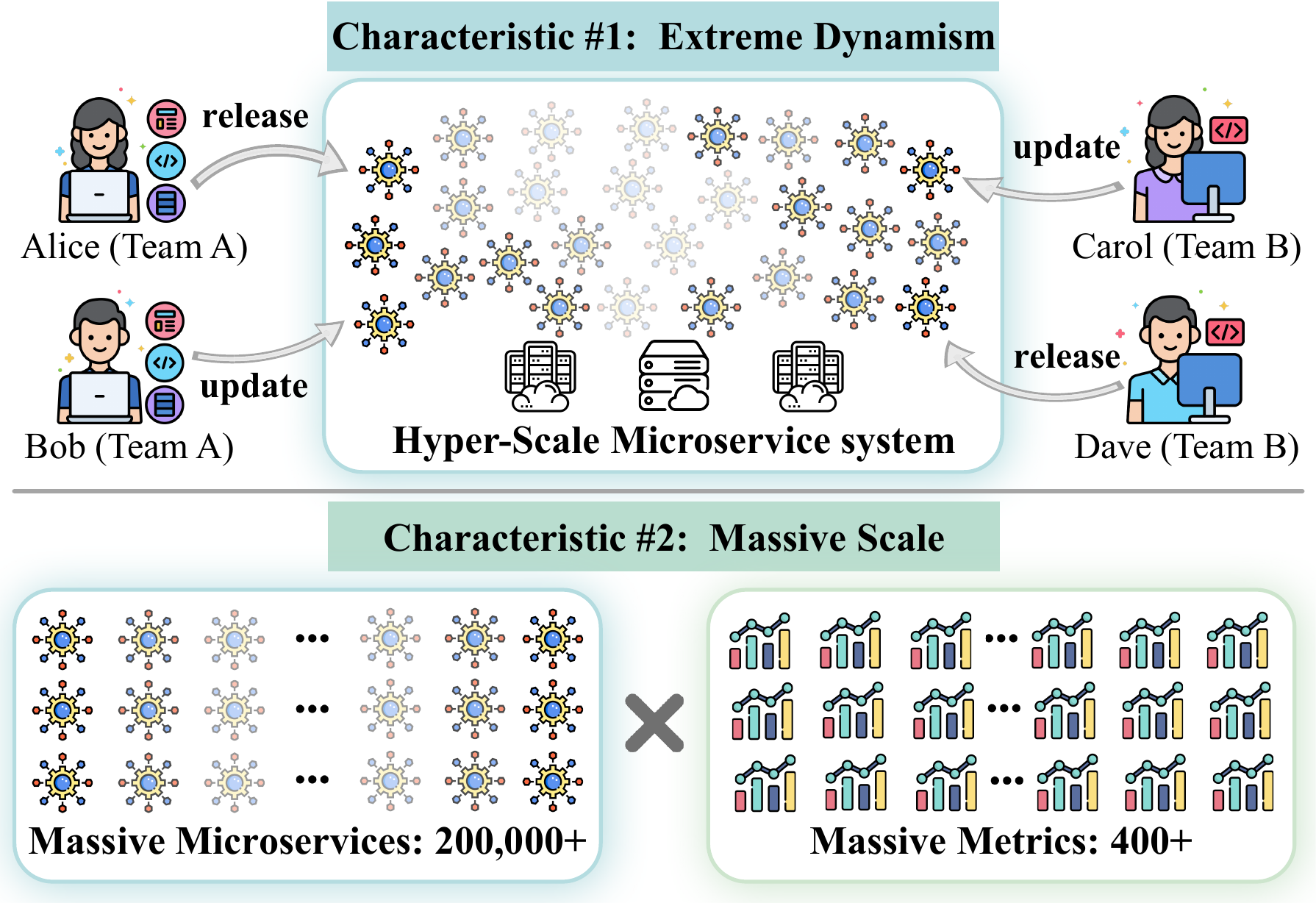}
    \caption{Two primary characteristics of hyper-scale microservice systems: extreme dynamism driven by decentralized, continuous updates across multiple teams, and massive scale involving hundreds of thousands of services monitored by numerous metrics.}
    \label{fig:intro}
    \vspace{-22pt}
\end{figure}

Microservice architectures have become the mainstream application framework for large-scale internet companies due to their inherent advantages, particularly their exceptional scalability~\cite{cui2025logeval,  pei2025flow,sun2024art, xie2026foundroot,xie2024microservice, zhang2025too, zhao2023robust}.
In a microservice architecture, a complex application is decomposed into a set of independently deployable and developmental services.
With continuous feature updates and rapid business growth, these architectures steadily burgeon into hyper-scale microservice systems.
For example, Kuaishou's hyper-scale microservice system already comprises over 200,000 microservices, ensuring 24$\times$7 availability to deliver entertainment for more than 400 million global users daily.
However, due to frequent service updates and massive scale expansion, system failures are inevitable.
RCA has thus become a critical technique in modern service operations ~\cite{ fu2025llm, li2023generic, wang2025kaiops, xu2025openrca, zhang2021robust, zhaollms, zhu2024hemirca}.
Its primary goal is to help SREs rapidly diagnose the origin of a failure, which involves two core tasks: 1) \textit{root cause service localization}---pinpointing the specific service responsible for the failure; and 2) \textit{failure type classification}---determining the type of the failure (\eg CPU saturation or MySQL slow query).
Accurate and efficient RCA can significantly reduce the Mean Time to Recovery (MTTR) and mitigate the impact of failures on end users.

Compared to open-source microservice benchmarks widely used in existing RCA literature—such as Google Online Boutique (10 services)~\cite{google_online_boutique} and even the largest industrial-scale study to date, which involves up to 30,000 services~\cite{liu2021microhecl}—performing RCA in hyper-scale microservice systems presents unique challenges characterized by two primary dimensions: extreme dynamism and massive scale, as illustrated in Fig.~\ref{fig:intro}. 
First, the system is highly dynamic. Microservices are developed and updated by different teams with completely independent release cycles. This decentralized governance means the topology and normal behavior patterns of the hyper-scale system are constantly evolving, demanding an RCA system that operates in real time without relying on pre-trained models.
Second, the scale of observability data is massive. Hyper-scale microservice systems encompass hundreds of thousands of microservices, each monitored by hundreds of metrics. When a failure occurs, it can quickly propagate through complex dependency chains and induce anomalies in a large number of downstream services and metrics. The combined growth in the number of services and the dimensionality of metrics therefore requires an RCA system with high efficiency.

Existing RCA approaches exhibit fundamental limitations when confronted with these two characteristics. On the one hand, deep learning-based methods \cite{li2021practical, liu2025causelens, sun2024art, xie2024microservice, yao2024sparserca, yu2023nezha} can efficiently localize failure once trained. However, they require frequent model retraining to capture the constantly changing system features of a hyper-scale system. In practice, it is difficult to define exactly when retraining is needed, and more importantly, the training process is so time-consuming that a single training cycle often takes longer than the system's variation cycle, rendering these models unusable for real-time deployment. 
On the other hand, recent Large Language Model (LLM)-based RCA approaches \cite{chen2024automatic, han2024potential, pei2025flow, roy2024exploring,sun2025trioxpert, wang2025towards, wang2024rcagent, zhou2023d} attempt to leverage in-context learning to accurately acquire new system features on the fly. However, in a hyper-scale system, crucial evidence of failure propagation is often buried within prohibitively long contexts composed of massive services and metrics. Consequently, LLMs easily fall into the ``Lost in the Middle'' dilemma \cite{liu2024lost}, failing to extract the true causal relationships and generating hallucinated diagnoses.

To bridge this gap, we propose \name \footnote{K is an acronym for the first letters of the Chinese words Kuaisu and Kuaishou, where Kuaisu means fast in Chinese.}, an end-to-end RCA system designed to efficiently localize root cause services and classify failure types in hyper-scale microservice systems.
Inspired by the principle of hierarchical RCA \cite{chen2014causeinfer}, \name follows a progressive multi-stage diagnosis paradigm that systematically narrows down the search space.
Specifically, the workflow is decomposed into three sequential steps: (1) Identifying the top-N most suspicious services from the massive dependency graph; (2) Constructing an initial causal graph for the anomalous metrics of each suspicious service; (3) Leveraging LLM-based reasoning to refine and complete the causal graph, and ultimately to localize the root cause service and classify the failure type.

To implement this concept into a practical system, three technical challenges must be addressed:

\textbf{Challenge 1:} A single failure case can involve a large number of services, and in extreme cases this number can exceed 10{,}000 in a hyper-scale system. This massive blast radius of downstream services makes it difficult to distinguish the true root cause service. 

\textbf{Challenge 2:} A single service in a hyper-scale system contains hundreds of metrics. 
Traditional causal discovery algorithms infer causal relationships directly from raw time-series data, overlooking the rich semantic information carried by the metrics. 
Our empirical study shows that, as the number of metrics increases, the accuracy of these algorithms declines sharply, while the computational cost becomes prohibitive.

\textbf{Challenge 3:} Although LLMs can use the semantic cues encoded in metric names and gather external evidence through tool calls to infer causal relationships, an efficient RCA system still needs stronger reasoning tailored to RCA scenarios. In particular, the LLM must integrate cross-domain knowledge (\eg CPU and MySQL) to verify causality, while avoiding the high latency caused by multi-turn reasoning.

\name addresses these challenges through three core design components. 
First, to tackle Challenge 1, \name employs an API-level drilldown strategy. Starting from the alerting Application Programming Interfaces (APIs), it recursively traverses the dependency graph with a robust scoring function, thereby pruning irrelevant downstream services and narrowing the search space to a compact set of suspicious services. 
Second, to address Challenge 2, \name utilizes Skeleton-based Causal Graph Instantiation. Instead of relying on time-consuming statistical algorithms, it leverages the semantic information of metrics to map anomalous time series into a generic causal skeleton graph with high recall. Third, to overcome Challenge 3, \name introduces a multi-agent reasoning framework with Retrieval Augmented Generation (RAG) and tiered memory, enabling specialized agents to collaboratively verify causality and gather evidence, while also accelerating the overall reasoning process through parallel execution, thereby enabling efficient LLM-based reasoning for RCA.
% We remark that the challenges we address here are not unique to Kuaishou but generally apply to other hyper-scale microservice systems, which ensures that \name can be extensively adopted in those systems as well. 

In summary, this paper makes three main contributions: (1) To the best of our knowledge, \name is the first end-to-end root cause analysis system designed for hyper-scale microservice systems. It can localize the root cause service and classify the failure type within a practical time budget for real-world incident response; (2) We propose a progressive multi-stage RCA framework that addresses several key challenges in hyper-scale systems. \name first combines API-level drilldown over multiple API-related signals to reduce the search space, structured causal constraints to organize anomalous metrics, and multi-agent reasoning to verify causal relations, forming a complete end-to-end analysis pipeline; (3) We evaluate \name on 300 real-world failures and deploy it in Kuaishou's production environment for over six months. The experimental results show that \name achieves AC@1 scores of 0.88 and 0.79 for root cause service localization and failure type classification, respectively, outperforming the strongest baseline by absolute gains of 31\% and 32\%, respectively. In production, \name reduces the average time for root cause localization by 77.3\%.

% We propose \name, an end-to-end root cause analysis system for hyper-scale microservice systems, which accurately localizes root cause services, classifies failure types, and generates diagnostic reports that support rapid troubleshooting by SREs, while maintaining a practical balance between diagnostic accuracy and the efficiency requirements of real-time production environments. We further propose a progressive multi-stage RCA framework that addresses key practical challenges in hyper-scale systems. Specifically, \name integrates API-level drilldown for search-space reduction, structured causal constraints for organizing anomalous metrics, and multi-agent reasoning for causal verification, thereby forming a robust end-to-end analysis pipeline. We deploy \name in the production environment of Kuaishou and evaluate it through more than six months of continuous operation. On average, \name processes more than 2,000 alerts per day, identifies root causes for 82\% of emergency incidents, and reduces diagnosis time from more than 30 minutes to less than 5 minutes.

%% file: Chapter/Motivation.tex
\subsection{Background}
\textbf{Hyper-scale microservice systems.} Hyper-scale microservice systems consist of hundreds of thousands of loosely coupled services that are independently developed and deployed. This immense scale is driven by fine-grained functional decomposition, as well as the proliferation of service versions introduced by canary releases and multi-region deployments. Through interactions based on Remote Procedure Call (RPC) and HTTP, these services form highly complex dependency graphs. 
Middle-platform services play a central role in such systems by providing shared capabilities, \eg traffic scheduling and risk control, to hundreds of upstream and downstream services. As a result, any failure that originates in, or propagates through, middle-platform services can rapidly expand the affected subgraph from dozens of services to thousands, thereby creating a massive blast radius of cascading anomalies.

% \textbf{Metrics.} To ensure system reliability, SREs rely on telemetry metrics continuously collected by monitoring platforms. A metric is a time-series measurement used to describe the state or behavior of a service over time, such as resource usage, runtime status, dependency health, and request performance. Different metrics may have different sampling granularities, ranging from fine-grained intervals such as 1 second to coarser intervals such as 10 minutes. Since these metrics originate from different components and operational aspects of a service, they reflect different facets of service behav   ior: some characterize external changes imposed on the service, such as request volume; some reflect its internal runtime state, such as CPU saturation; some describe the status of downstream dependencies or middleware, such as MySQL latency; and others directly indicate service performance, such as success rate or response latency.

\subsection{Motivation}
\label{sec:motivation}
To motivate the design of our system, we conducted an empirical study based on production incidents in hyper-scale microservice systems.

\textbf{Motivation 1:} Service-level observability is too coarse for RCA in hyper-scale microservice systems.

\begin{figure}[t]
    \centering
    \includegraphics[width=\linewidth]{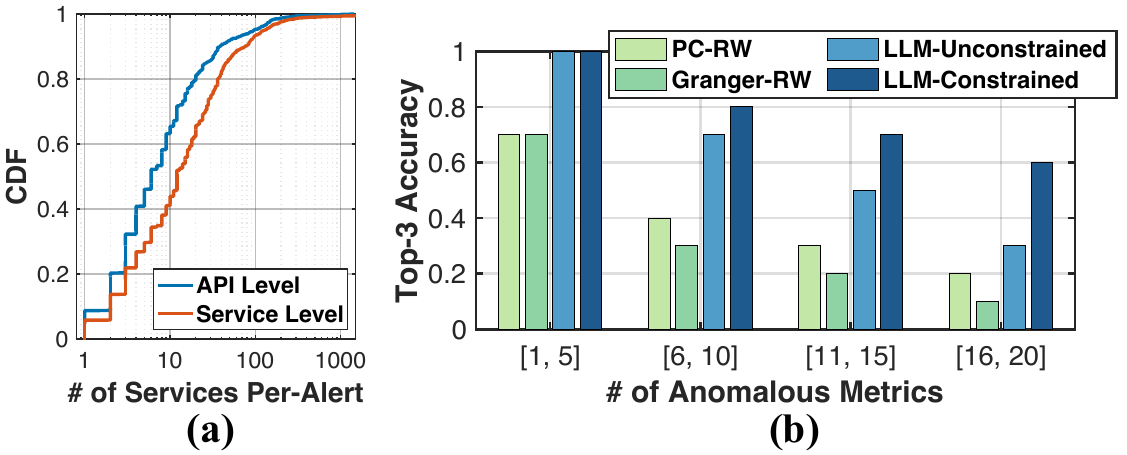}
    \caption{Empirical study on the limitations of existing RCA practices. (a) CDF of the number of services from the alerting service to the root cause service under API-level and service-level dependencies. (b) Top-3 root cause metric localization accuracy under different numbers of anomalous metrics.}
    \label{fig:motivation}
    \vspace{-12pt}
\end{figure}

In production environments, service-level metrics are often used as the default granularity for RCA \cite{liu2021microhecl, yang2025hg, zhang2023robust}. However, this granularity is too coarse for accurate root cause service localization, because it aggregates multiple APIs within a service and therefore obscures the fine-grained failure signals associated with individual invocation paths. 
On the one hand, service-level diagnosis mixes multiple invocation paths from different APIs within the same service, which causes all downstream services related to that service to be indiscriminately included in the RCA search space.
As shown in Fig.~\ref{fig:motivation}(a), service-level dependencies consistently involve more candidate services than API-level dependencies. 
Specifically, 36.3\% of alerts involve more than 10 services under API-level dependencies, whereas this proportion increases to 56.7\% under service-level dependencies. On the other hand, the service-level success rate is dominated by high-traffic APIs, which can mask severe degradation in low-traffic but faulty APIs \cite{liu2025causelens}. For example, within the same service, API\#1 handles 680K Queries Per Second (QPS), whereas API\#2 handles only 1.6K QPS. Even when the success rate of API\#2 drops sharply to 10\%, the aggregated service-level success rate remains above 99\% because the traffic volume of API\#1 overwhelmingly dominates the overall service-level metric. As a result, the failure of the faulty API remains hidden at the service level.

\textbf{Design insight 1:} Root cause analysis in hyper-scale microservice systems should begin with API-level metrics rather than service-level metrics.
\begin{figure*}[t]
    \begin{minipage}{0.7\textwidth}
        \centering
        \includegraphics[width=\linewidth]{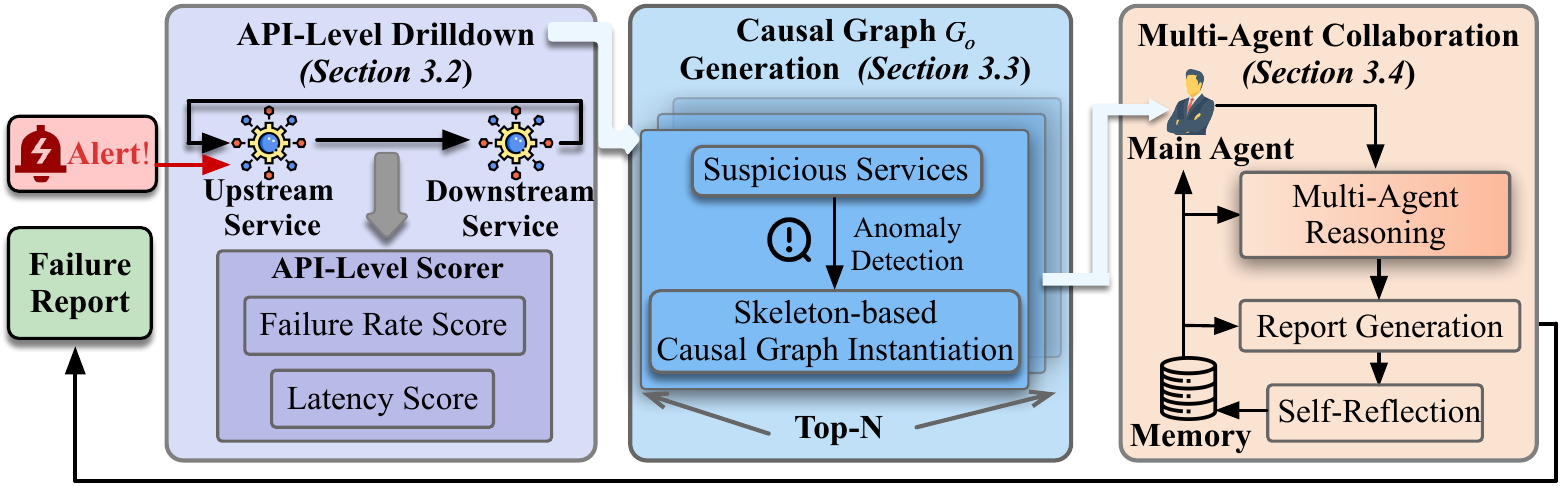} 
        \caption{System architecture of \name. The workflow consists of three modules: (1) API-Level Drilldown to locate the top-$N$ suspicious services from an alert; (2) Causal Graph $G_o$ Generation to construct an initial skeleton-based causal graph for anomalous metrics; and (3) Multi-Agent Collaboration to refine the graph and generate the failure report.}
        \label{fig:overview}
    \end{minipage}
    \hfill 
    \begin{minipage}{0.29\textwidth}
        \centering
        \includegraphics[width=\linewidth]{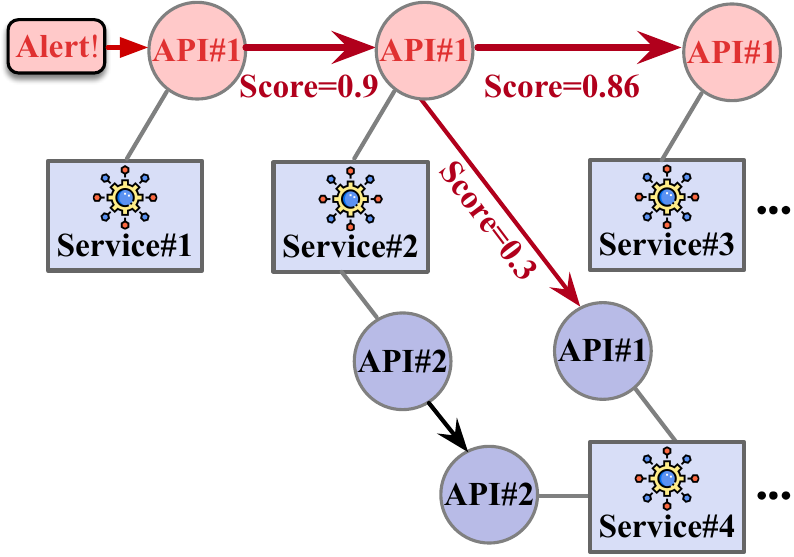} 
       \caption{Illustration of the API-level drilldown process. Starting from the alerting API, \name recursively evaluates and prunes downstream APIs based on a scoring function.}
        \label{fig:api_level}
    \end{minipage}
    \vspace{-1em}
\end{figure*}
\textbf{Motivation 2:} Causality should not be inferred directly from the raw anomalous metrics set.

After locating root cause service, a common RCA process is to detect anomalous metrics, infer causality among them using causal discovery algorithms, and then apply graph-based ranking methods such as Random Walk (RW) \cite{lawler2010random} to locate root cause metric.
Prior work \cite{chen2014causeinfer, lin2024root, meng2020localizing, zhang2024illuminating} typically relies on traditional causal discovery algorithms, such as Peter-Clark (PC) \cite{spirtes2000causation} and Granger causality \cite{granger1980testing}, to construct causality among metrics in the anomalous metric set. 
However, production metric data are often high-dimensional (typically more than 10 metrics) and noisy due to sampling distortion and discontinuities, making these time-series-based causal discovery methods unreliable in production environments.
As shown in Fig.~\ref{fig:motivation}(b), we collected 20 real-world incidents from Kuaishou with varying numbers of anomalous metrics, and applied the PC and Granger algorithms to each case. The results show that the performance of both PC and Granger degrades sharply as the number of metrics increases. 
When only 5 metrics are involved, both methods achieve an accuracy of around 50\%. However, when the number of metrics reaches 20, which is a common scale in hyper-scale systems, the accuracy of both methods falls below 20\%. Moreover, as the number of anomalous metrics and the length of the time series increase, the computational cost of traditional causal discovery becomes too high for real-time RCA in hyper-scale systems.

Recent studies have shown that LLMs possess certain causal reasoning capabilities: they can recognize variable semantics and assist causal discovery by incorporating prior knowledge \cite{du2025causal, feng2025reliability, wan2024large}. 
As shown in Fig.~\ref{fig:motivation}(b), we further evaluate LLM-based causal discovery under both unconstrained and constrained settings. When the number of metrics is small, the unconstrained LLM achieves high accuracy, reaching 100\%. However, as the number of anomalous metrics increases, its accuracy also drops rapidly to around 30\%. 
This decline occurs because LLMs tend to be overconfident and hallucinate non-existent causal relations when they reason directly over a large set of anomalous metrics. A more robust approach is to impose structured causal constraints on the anomalous metric set using expert knowledge before causal reasoning. For example, directional constraints restrict candidate causal edges to those that are consistent with realistic failure propagation, whereas semantic constraints specify that certain availability-related metrics can only be treated as effects rather than causes. Under these constraints, the LLM remains substantially more robust, and its accuracy stays above 60\% even as the number of anomalous metrics increases.

\textbf{Design insight 2:} Anomalous metrics should first be organized into a sparse, high-recall structural prior that constrains the causal search space, after which LLMs can perform targeted, evidence-based reasoning to refine the final causal explanation.

%% file: Chapter/SystemDesign.tex
\subsection{Overview}
\name is mainly composed of three modules, as shown in Fig.~\ref{fig:overview}. (1) \textbf{API-level drilldown module.} Starting from the alerting API, this module recursively traverses the service dependency graph and scores downstream APIs using failure rate and latency. It keeps only suspicious propagation paths, reducing the graph to a compact set of suspicious services; (2) \textbf{Causal graph $G_o$ generation module.} For each suspicious service, this module detects anomalous metrics from monitoring data and maps them onto a generic causal skeleton with four meta-metric types. This process yields an initial service-level causal graph $G_o$, which provides structured prior knowledge for later diagnosis; (3) \textbf{Multi-agent collaboration module.} This module further refines $G_o$ and improves diagnostic accuracy through a multi-agent framework in which a Main Agent coordinates multiple Sub Agents to verify causal relations and gather evidence. A memory-augmented retrieval mechanism supplies similar historical cases and diagnostic experience. After several rounds of refinement, the final causal graph is used to generate a failure report that identifies the root cause service and failure type.

\subsection{API-level drilldown}
\label{sec: where}

In our hyper-scale microservices environment, the initial signal of a failure is often an availability alert on a specific service API, triggered when its success rate drops below a predefined threshold. 
To rapidly pinpoint the cause of this availability drop, \name must first drilldown from a vast number of downstream dependencies to a small set of suspicious services. 
The primary challenge lies in selecting accurate and efficient features to guide this drilldown process.

% Drawing on established SRE principles from Google \cite{beyer2016site}, service health can be characterized by four golden metrics: latency, traffic, errors, and saturation.
% Among them, latency and errors (\ie failure rate) provide the most direct signals of fault propagation.
% A fault in a downstream service typically propagates through the dependency graph and appears as increased errors or latency in its upstream caller.
% By contrast, anomalies in traffic or saturation within a downstream service do not necessarily affect upstream callers, because services differ in capacity and resilience mechanisms.
% Therefore, following this principle and the observations in Section~\ref{sec:motivation}, we focus drilldown on failure rate and latency at the API level.

Following established SRE principles~\cite{beyer2016site} and the observations in Section~\ref{sec:motivation}, we focus drilldown on failure rate and latency at the API level, as they provide the most direct signals of fault propagation: a fault in a downstream service typically manifests as increased errors or latency in its upstream caller. We quantify the likelihood that a failure observed at an upstream API $P$ (Parent) originates from a downstream API $C$ (Child) with the following scoring function:
\begin{equation}
\text{Score}(P,C) = \max (\text{Score}_{f}(P,C), \text{Score}_{l}(P,C))
\label{eq:score_total}
\end{equation}
where $\text{Score}_{f}(P,C)$ is the failure rate score and $\text{Score}_{l}(P,C)$ is the latency score. 
% The use of the $\max$ function is a deliberate design choice. 
% On one hand, it allows the most severe anomaly to dominate the final score, preventing its significance from being diluted by the normal behavior of the other metric (\eg when failure rate spikes dramatically but latency remains stable). 
% On the other hand, the max operation is computationally efficient, which is critical for real-time analysis in a hyper-scale system. 
If $\text{Score}(P,C)$ exceeds threshold, we regard the downstream service as suspicious and continue the drilldown from that point.

\textbf{Failure rate score.} 
Failures in a downstream API $C$ often propagate to its upstream caller API $P$, leading to highly synchronized fluctuation patterns in their failure rate time series.
To quantify this temporal dependency, we define the failure rate score as follows:
\begin{equation}
\text{Score}_{f}(P, C) = \max_{\tau \in [0, L]} \text{Corr}\left( \mathbf{p}[t_s, t_e], \mathbf{c}[t_s - \tau, t_e - \tau] \right)
\label{eq:score_fail}
\end{equation}
where $\mathbf{p}$ and $\mathbf{c}$ denote the failure rate time series of the upstream API $P$ and downstream API $C$, respectively. 
$L$ represents the maximum time lag.
The correlation is computed using the \textit{Pearson Correlation Coefficient} \cite{benesty2009pearson} over a dynamic observation window $[t_s, t_e]$, where $t_e$ denotes the timestamp at which the alert is triggered.
To concentrate on the failure period, we determine the starting point $t_s$ as the inflection point, defined as the latest moment before $t_e$ at which the sign of the first-order difference in the failure rate changes, indicating the qualitative transition from a steady state to a failure state.
% Given that our system uses a 1-minute sampling interval and that failure propagation typically remains within 5 minutes, we set the maximum time lag to $L = 5$.
Finally, we perform a $p$-value test with $\alpha = 0.05$ to ensure statistical significance; correlations that do not pass this test are set to zero to filter out spurious noise.
\begin{figure}[!t]
    \centering
    \captionsetup[subfigure]{font=small}
    \begin{subfigure}[]{0.35\textwidth}
        \centering
        \includegraphics[width=\textwidth]{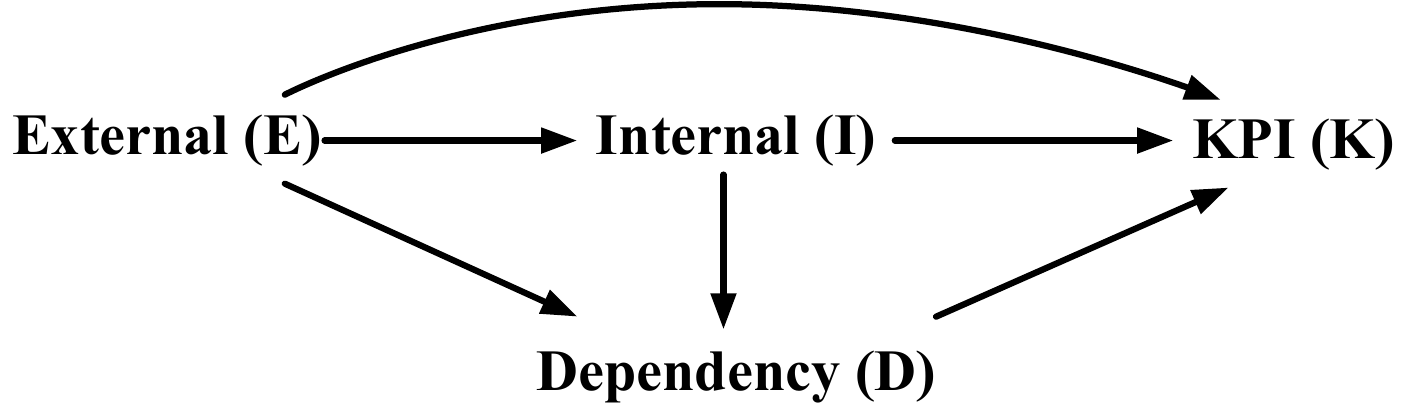}
        \caption{}
        \label{fig:causal_graph1}
    \end{subfigure}%
    \hfill
    \begin{subfigure}[]{0.4\textwidth}
        \centering
            \includegraphics[width=\textwidth]{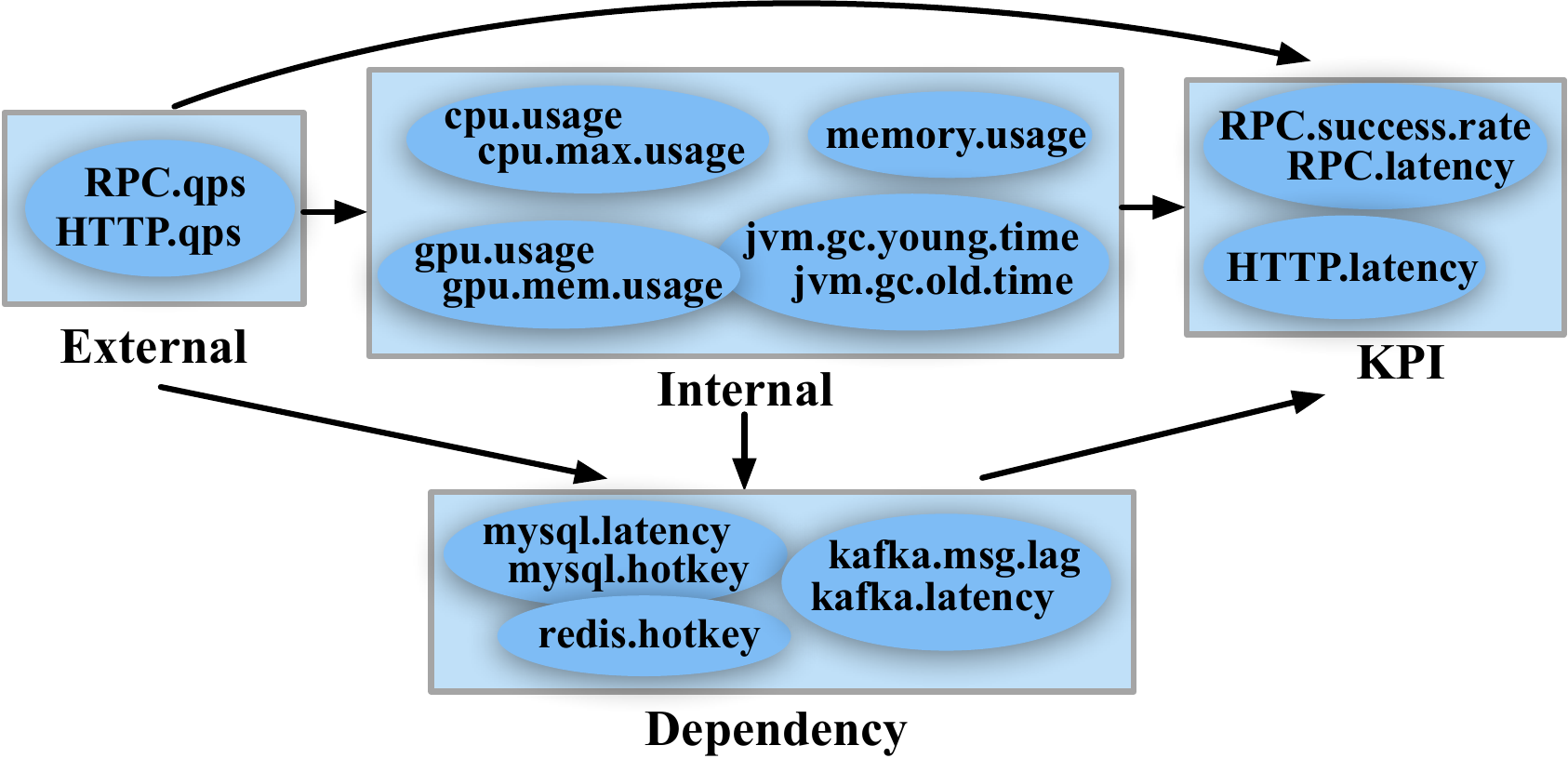}
        \caption{}
        \label{fig:causal_graph2}
    \end{subfigure}
    \caption{Skeleton-based causal graph instantiation. (a) The generic causal skeleton graph $G_s$ defining the predefined causal relationships among four meta-metric types. (b) An instantiated causal graph example where anomalous metrics (\eg cpu.usage, mysql.latency) are mapped to their corresponding meta-metric types based on the skeleton structure.}
    \label{fig:causal_graph}
    \vspace{-1em}
\end{figure}

\textbf{Latency score.} Unlike the failure rate, latency is influenced by multiple factors (\eg network jitter and code execution), which leads to a lower signal-to-noise ratio.
Consequently, relying solely on correlation often produces inaccurate results.
To address this, we define a composite score that integrates anomaly degree, fluctuation contribution, and correlation degree:
\begin{equation}
\text{Score}_{l}(P, C) = w_1 \cdot \mathcal{A}(C) + w_2 \cdot \mathcal{F}(P, C) + w_3 \cdot \mathcal{C}(P, C)
\label{eq:score_latency}
\end{equation}
where $w_1$, $w_2$, and $w_3$ are balancing weights.
All components are computed over the same dynamic observation window $[t_s, t_e]$ defined in the failure rate.
Let $\mathbf{p}$ and $\mathbf{c}$ represent the latency time series of the upstream API $P$ and downstream API $C$, respectively, and let $p_t$ and $c_t$ denote their values at time $t$. 
Firstly, the anomaly degree $\mathcal{A}$ quantifies the average deviation of the downstream API $C$ from its optimal performance baseline within the observation window:
\begin{equation}
\mathcal{A}(C) = \frac{1}{t_e - t_s + 1} \sum_{t=t_s}^{t_e} \frac{|c_t - c_{base,t}|}{c_t}
\end{equation}
where $c_{base,t}$ denotes the historical minimum latency for the corresponding observation window, calculated across a multi-granularity look-back (1-hour, 1-day, and 1-week).
Secondly, the fluctuation contribution $\mathcal{F}$ measures the extent to which the latency instability of the downstream API $C$ explains the fluctuation of its caller $P$:
\begin{equation}
\mathcal{F}(P, C) = \frac{\text{QPS}_C}{\text{QPS}_P} \cdot \frac{\sum_{t=t_s}^{t_e} |c_t - c_{t-1}|}{\sum_{t=t_s}^{t_e} |p_t - p_{t-1}|}
\end{equation}
where QPS denotes the Queries Per Second of the corresponding APIs.
Thirdly, the correlation degree $\mathcal{C}$ is used to validate trend consistency. It adopts the same time-lagged Pearson Correlation Coefficient defined in Eq.~\ref{eq:score_fail}.

The overall drilldown process is a recursive traversal of the service dependency graph guided by $\text{Score}(P,C)$. 
As depicted in Fig.~\ref{fig:api_level}, this process begins at the alerting API (Service\#1::API\#1).
The drilldown process first evaluates the direct downstream call to Service\#2::API\#1. 
Since the score of 0.9 exceeds the threshold, the process continues down this path. 
From Service\#2::API\#1, the process then examines its own downstream dependencies. 
The path leading to Service\#3::API\#1 yields a high score of 0.86, which prompts the process to continue along this branch. 
Conversely, the path to Service\#4::API\#1 is pruned because its score of 0.3 falls below the threshold. This indicates that Service\#4 is not a major factor in the performance degradation.
The drilldown process repeats throughout the dependency graph until no further paths meet the criteria. Through this process, \name effectively narrows the candidates to a compact set of suspicious APIs. Finally, all identified APIs are ranked based on their calculated scores. 
The top-$N$ services associated with the highest-scoring APIs are recommended as the most suspicious root causes for the alert.
\begin{figure*}[!t]
    \centering
    \includegraphics[width=\linewidth]{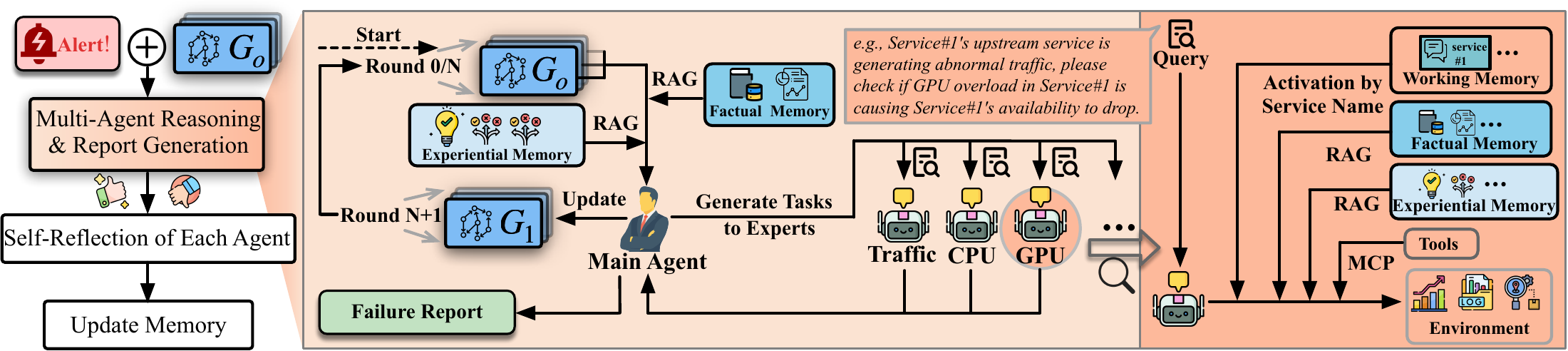} 
    \caption{Overview of the Multi-Agent Collaboration framework. The Main Agent orchestrates domain-specific Sub Agents (\eg Traffic, CPU, GPU) to iteratively verify and refine the initial causal graph $G_o$. A three-tiered memory system (Working, Factual, and Experiential Memory) retrieves historical failure patterns via RAG to support evidence-based reasoning and continuous learning through self-reflection.}
    \label{fig:mas}
    \vspace{-1em}
\end{figure*}
\subsection{Skeleton-based Causal Graph Instantiation}
\label{sec: what}

In Section~\ref{sec: where}, \name has narrowed down the scope to a few suspicious services.
However, in hyper-scale systems, a single service is typically monitored by a vast array of metrics—often exceeding 400. 
These metrics cover physical resources (\eg CPU and GPU), dependent middleware and database (\eg Kafka and MySQL) and service availability Key Performance Indicator (KPI). 
To effectively diagnose an alert, it is crucial to discover the causal relationships among these metrics.
A straightforward approach is to apply standard causal discovery algorithms (\eg PC) to the anomalous metrics.
However, as demonstrated in our empirical study (Section~\ref{sec:motivation}), the accuracy of these algorithms drops sharply as the number of metrics increases, while the computational overhead becomes prohibitive.

Therefore, instead of attempting to recover a complete causal graph directly from raw time-series, the objective of this section is to construct an initial causal graph, denoted as $G_o$, within each suspicious service.
The rationale is to provide structured prior knowledge of the failure for the subsequent multi-agent refinement.
Intuitively, this mirrors the conceptual design of Bayesian inference \cite{box2011bayesian}, offering a principled and systematic diagnostic pipeline.

Inspired by CIRCA \cite{li2022causal} and cloud-native architectures, we propose a generic causal skeleton for individual services, denoted as $G_s$, which is composed of four types of meta-metrics: External (\textit{E}), Internal (\textit{I}), Dependency (\textit{D}), and KPI (\textit{K}), as illustrated in Fig.~\ref{fig:causal_graph}(a). The definitions of these meta-metrics and their relationships are described as follows: (1) \textbf{External (\textit{E})}: \textit{E} represents external factors, such as traffic surges or upstream poison requests. Anomalies in \textit{E} can directly impair \textit{K}, induce anomalies in \textit{I} by increasing processing load, or overload \textit{D} through high-frequency access;
(2) \textbf{Internal (\textit{I})}: \textit{I} represents host-level or component-specific resources, such as CPU or GPU utilization. Although \textit{I} directly affects \textit{K}, anomalies in \textit{I} may also degrade the performance of \textit{D} through abnormal behaviors, such as aggressive retries or prolonged connection occupancy;
(3) \textbf{Dependency (\textit{D})}: \textit{D} represents the external dependencies of a service, such as Kafka, MySQL, and Redis. Anomalies in \textit{D} can directly impair \textit{K};
(4) \textbf{KPI (\textit{K})}: \textit{K} represents the key performance indicators of a service, such as latency and success rate.

% Notably, in production environments, robust protection mechanisms (\eg Timeouts, Circuit Breaker, and Fallback) are typically deployed between a service and its dependencies. 
% These mechanisms prevent dependency issues from directly exhausting internal resources of the service (\eg CPU/GPU). Therefore, we do not model a direct causal relation from \textit{D} to \textit{I} in $G_s$.

Next, we use the skeleton graph $G_{s}$ to build a concrete intra-service initial causal graph $G_o$. 
The construction proceeds in two stages. 
We first detect anomalous metrics using a hybrid strategy to improve robustness across different metric types and services. We use Isolation Forest \cite{liu2008isolation} as the default detector because it does not depend on strong distributional assumptions and is more reliable for heterogeneous metrics collected from different services. In practice, however, we found that Isolation Forest can miss some shape-related anomalies, such as sustained shifts and plateau-like degradations. To address this issue, we combine it with PatternMatcher \cite{wu2021identifying}, which is better at capturing temporal anomaly patterns that are important in production.
We then map each detected anomalous metric to its meta-metric type and instantiate causal edges among these meta-metrics according to the predefined structure of $G_{s}$. Fig.~\ref{fig:causal_graph}(b) shows an example. A sudden traffic surge (E) increases internal pressure (I), which appears in metrics such as cpu.usage and gpu.usage, and this change further affects RPC.latency and RPC.success.rate (K). Meanwhile, the increase in internal pressure also triggers anomalies in dependencies (D), such as mysql, redis, and kafka.

However, the causal graph $G_o$ instantiated from the generic skeleton $G_{s}$ remains insufficient to capture the full complexity of failure propagation within a service. 
For example, in Fig.~\ref{fig:causal_graph}(b), the anomalies of mysql.hotkey and redis.hotkey are caused by the traffic surge rather than by internal components of the service, such as GPU.
Furthermore, causal relations may exist within a single meta-metric type, which cannot be represented by $G_{s}$. 
To address this, the next section introduces a multi-agent system to validate the causal relations in $G_o$ and discover the finer-grained causal relations among metrics within the same meta-metric type, thereby completing the causal graph.

\subsection{Multi-Agent Collaboration}
\label{sec: which}

Following the process described in the previous section, \name identifies a set of suspicious services and constructs an initial causal graph $G_o$ for each service. 
However, these graphs often contain redundant causal edges and fail to capture finer-grained causal relations across meta-metrics. 
The final challenge is therefore to refine these graphs accurately by pruning spurious edges and uncovering implicit causal relations.

A naive solution is to present all causal edges from all suspicious services directly to a single LLM and ask it to assess their validity. 
However, this approach is prone to a critical form of hallucination. Because the skeleton graph is generally plausible, the LLM tends to accept the causal relations within each service as valid. 
Our key insight is that this ambiguity can be reduced by exploiting the rich semantic information encoded in metric names (\eg cpu.max.usage, mysql.slow.count) together with broader cross-service context. 
Based on this insight, we design a multi-agent architecture that separates high-level coordination from low-level causality verification, as illustrated in Fig.~\ref{fig:mas}.

\textbf{Agent definition.}
\name adopts a \textit{master-slave} multi-agent architecture in which a single \textit{Main Agent} serves as the central orchestrator and coordinates a pool of specialized \textit{Sub Agents}. 
As described in Section~\ref{sec: what}, anomalous metrics are first mapped to broad meta-metric categories (\textit{E}, \textit{I}, \textit{D}, \textit{K}). 
Each meta-metric category covers multiple finer-grained failure scenarios. 
For example, an anomaly in the internal meta-metric may arise from a CPU bottleneck or a memory leak, and these cases require different diagnostic knowledge and tools. 
Accordingly, we define nine domain-specific Sub Agents aligned with classic failure scenarios (\eg Traffic Expert, CPU Expert, and GPU Expert). 
Each Sub Agent is equipped with a domain-specific prompt, common query tools, and specialized diagnostic tools (\eg SQL analyzer). 
The Main Agent maintains a global view of all $G_o$ graphs and dispatches causality verification tasks to the appropriate Sub Agents. These Sub Agents collect concrete evidence to confirm or refute localized causal links, which enables the pruning of spurious edges and the discovery of implicit intra-meta-metric causality.
% Each Sub Agent is equipped with three components: (1) a system prompt encoding its domain expertise for its specific failure scenario; (2) a set of common tools for universal tasks like querying metric values; and (3) specialized tools tailored to its domain (\eg a SQL analysis tool for the MySQL Expert).
% Specifically, the Main Agent maintains a global view of the initial causal graphs $G_o$ from all suspicious services. 
% It leverages its understanding of metric semantics and cross-service relationships to dispatch a series of causality verification tasks to the corresponding Sub Agents, whose responsibility is to gather concrete evidence (\eg querying metric values, fetching logs, checking Java thread pools) to confirm or refute a specific, localized causal link. 
% This process not only prunes spurious inter-meta-metric causal edges proposed by the skeleton graph but also enables the discovery of finer-grained, intra-meta-metric causal edges that were absent in the initial graph.

\textbf{Memory system.} To preserve reasoning continuity and exploit accumulated experience, \name incorporates a three-tier memory system. First, \textit{Working Memory} records the short-term dialogue history for each service. It supports \textit{cross-service activation}, which provides holistic context when a task involves multiple services. For long-term knowledge, we separate episodic cases from transferable insights to avoid retrieval interference. \textit{Factual Memory} stores service-specific historical failures, enabling in-context learning from empirical evidence and thereby reducing hallucinations. \textit{Experiential Memory} stores generalized reasoning insights (\eg ``traffic saturation causes MySQL slow queries''), allowing the LLM to make informed judgments even for previously unseen services.

\textbf{Retrieval-Augmented Generation.} 
\name employs a RAG mechanism to incorporate historical knowledge dynamically. At each reasoning step, the module retrieves the top-$3$ relevant entries, prioritizing Factual Memory and then falling back to Experiential Memory. The Main Agent issues queries using the full alert context, whereas Sub Agents use task-specific metrics. The retrieval mechanism uses a hybrid scoring model that combines lexical similarity with temporal relevance. Given a query, the similarity score for a historical entry is computed as:
\begin{equation}
\text{Score}(Q, H) = \alpha \cdot S(Q_{s}, H_{s}) + \beta \cdot S(Q_{m}, H_{m}) + \gamma \cdot e^{-\lambda (T_Q - T_H)}
\label{eq:rag}
\end{equation}
where $S(\cdot, \cdot)$ denotes BM25-based lexical similarity~\cite{robertson2004simple}, computed over service names (subscript $s$) and anomalous metric names (subscript $m$), respectively, while the exponential term penalizes older entries. The weights $\alpha$, $\beta$, and $\gamma$ are tuned separately for each memory type to reflect their distinct retrieval objectives: Factual Memory prioritizes service names and recency to retrieve recent incidents from the same service, whereas Experiential Memory emphasizes metric names to retrieve transferable insights across services.

\textbf{Putting it together.} 
The collaboration begins with the Main Agent performing a global RAG query and decomposing graph refinement into parallel tasks. At the same time, Sub Agents perform focused RAG queries and use specialized tools to gather evidence. The Main Agent then synthesizes these findings to prune refuted edges and dispatches new tasks to identify missing intra-meta-metric links. Finally, the Main Agent generates a comprehensive failure report. The system then enters a self-reflection phase, during which case-specific details and generalizable insights are distilled into the memory system, enabling continuous evolution after each resolved alert.

%% file: Chapter/Evaluation.tex
In this section, we evaluate the performance of \name with the aim of answering the following questions:

\textbf{RQ1:} How does \name compare with existing methods?

\textbf{RQ2:} Does each component contribute to \name ? 

\textbf{RQ3:} How sensitive is \name to key hyperparameters?

\textbf{RQ4:} How beneficial is \name in assisting SREs with RCA?

\subsection{Experimental Setup}
\textbf{Dataset.} To comprehensively evaluate the performance of \name, we constructed a real-world failure dataset from the monitoring platform of Kuaishou's hyper-scale microservice system. The dataset contains 300 system-related failures that were sampled from October 2025 to March 2026, spanning multiple business lines (\eg Short Video, Recommendation, and E-commerce). In total, the dataset encompasses nine classical failure types commonly observed in hyper-scale microservice systems, including but not limited to abnormal traffic, CPU/GPU saturation and MySQL slow queries, thereby ensuring sufficient diversity for a robust evaluation. The ground truth for each failure, comprising both the root cause service and the failure type, was labeled by a dedicated annotation team consisting of five SREs with over five years of operational experience.

\textbf{Evaluation metrics.} We evaluate \name from three perspectives. For \textit{accuracy}, we adopt AC@k and Avg@k, two metrics widely used in root cause analysis, computed separately for root cause service localization and failure type classification, where AC@k measures whether the correct answer appears in the top-$k$ predictions, and $Avg@k=\frac{1}{k}\sum_{i=1}^{k} AC@i$. For \textit{efficiency}, we measure end-to-end latency from alert triggering to the generation of the failure report. For \textit{report usability}, we conduct a human-centered evaluation following~\cite{liu2024logprompt, zhaollms}, inviting SREs to rate generated reports along two dimensions: \textit{Readability} (clarity, logical organization, and technical fluency) and \textit{Usefulness} (the extent to which the diagnosis supports timely failure resolution).

% \textbf{Evaluation metrics.} We evaluate the performance of \name from three complementary perspectives: accuracy, efficiency, and report usability.
% For accuracy, we adopt AC@k and Avg@k, two metrics widely used in root cause analysis studies. Specifically, AC@k measures whether the ground-truth result appears within the top-$k$ candidates for a given case. Avg@k summarizes performance by averaging AC@k over all evaluated cases. Both metrics are computed separately for root cause service localization and failure type classification.
% For efficiency, we measure end-to-end latency, defined as the time from the alert being triggered to the generation of the final failure report.
% For report usability, conventional metrics based on lexical or semantic overlap often fail to capture practical value in operational settings. We therefore conduct a human-centered evaluation following the methodology in \cite{liu2024logprompt, zhaollms}. Specifically, we invited SREs to assess the failure reports generated with the assistance of \name along two dimensions: (1) \textit{Readability:} the clarity, logical organization, and technical fluency of the report; (2) \textit{Usefulness:} the extent to which the diagnosis supports timely failure resolution.
\begin{table*}[!t]
    \centering
    % \small
    \setlength{\tabcolsep}{3.8pt}
    \renewcommand{\arraystretch}{1.10}
    \caption{Comparing \name with Existing Methods on the Sub-task of RCA.}
    \label{tab:overall_exp}
    \vspace{0.3em}
    \begin{tabular}{cc@{\hspace{0.5em}}c@{\hspace{0.9em}}ccc@{\hspace{0.9em}}ccc@{\hspace{0.6em}}c}
        \toprule
        \multirow{2.5}{*}{\shortstack[c]{\textbf{Main Seed LLM}}}
        & \multirow{2.5}{*}{\shortstack[c]{\textbf{Sub Seed LLM}}}
        & \multirow{2.5}{*}{\textbf{Method}}
        & \multicolumn{3}{c}{\textbf{Root Cause Service}}
        & \multicolumn{3}{c}{\textbf{Failure Type}}
        & \multirow{2.5}{*}{\shortstack[c]{\textbf{Latency (s/case)}}} \\
        \cmidrule(lr){4-6} \cmidrule(lr){7-9}
        & &
        & \textbf{AC@1} & \textbf{AC@3} & \textbf{Avg@3}
        & \textbf{AC@1} & \textbf{AC@3} & \textbf{Avg@3} & \\
        \midrule
        
        \textbf{Claude-opus-4.6} & \textbf{GPT-5.1} & \textit{\name}
            & \textbf{0.88} & \textbf{0.96} & \textbf{0.92}
            & \textbf{0.79} & \textbf{0.98} & \textbf{0.89} & \textbf{146.59} \\
        \multicolumn{2}{c}{\textbf{Claude-opus-4.6}} & CoT \cite{wei2022chain}
            & 0.14 & 0.17 & 0.16
            & 0.12 & 0.34 & 0.24 & 32.87 \\
        \multicolumn{2}{c}{\textbf{Claude-opus-4.6}} & ReAct \cite{yao2022react}
            & 0.42 & 0.47 & 0.45
            & 0.29 & 0.44 & 0.37 & 211.48 \\
        \multicolumn{2}{c}{\textbf{Claude-opus-4.6}} & Reflexion \cite{shinn2023reflexion}
            & 0.48 & 0.52 & 0.50
            & 0.37 & 0.57 & 0.48 & 411.80 \\
        \multicolumn{2}{c}{\textbf{Claude-opus-4.6}} & RCA-Agent \cite{xu2025openrca}
            & 0.57 & 0.62 & 0.60
            & 0.47 & 0.72 & 0.61 & 342.80 \\
        
        \midrule
        
        \textbf{GPT-5.2} & \textbf{GPT-5.1} & \textit{\name}
            & \textbf{0.77} & \textbf{0.89} & \textbf{0.84}
            & \textbf{0.66} & \textbf{0.88} & \textbf{0.78} & \textbf{129.36} \\
        \multicolumn{2}{c}{\textbf{GPT-5.2}} & CoT
            & 0.17 & 0.22 & 0.20
            & 0.12 & 0.31 & 0.22 & 33.54 \\
        \multicolumn{2}{c}{\textbf{GPT-5.2}} & ReAct
            & 0.41 & 0.49 & 0.45
            & 0.29 & 0.45 & 0.38 & 166.82 \\
        \multicolumn{2}{c}{\textbf{GPT-5.2}} & Reflexion
            & 0.47 & 0.55 & 0.51
            & 0.35 & 0.53 & 0.45 & 298.47 \\
        \multicolumn{2}{c}{\textbf{GPT-5.2}} & RCA-Agent
            & 0.55 & 0.64 & 0.60
            & 0.44 & 0.67 & 0.57 & 255.91 \\
        \midrule
        \textbf{MiniMax-M2.7} & \textbf{DeepSeek-V3} & \textit{\name}
            & \textbf{0.70} & \textbf{0.85} & \textbf{0.78}
            & \textbf{0.60} & \textbf{0.81} & \textbf{0.72} & \textbf{95.74} \\
        \multicolumn{2}{c}{\textbf{MiniMax-M2.7}} & CoT
            & 0.14 & 0.19 & 0.17
            & 0.11 & 0.26 & 0.19 & 15.42 \\
        \multicolumn{2}{c}{\textbf{MiniMax-M2.7}} & ReAct
            & 0.33 & 0.40 & 0.37
            & 0.22 & 0.36 & 0.30 & 76.88 \\
        \multicolumn{2}{c}{\textbf{MiniMax-M2.7}} & Reflexion
            & 0.38 & 0.46 & 0.42
            & 0.27 & 0.43 & 0.36 & 136.91 \\
        \multicolumn{2}{c}{\textbf{MiniMax-M2.7}} & RCA-Agent
            & 0.46 & 0.56 & 0.52
            & 0.35 & 0.56 & 0.47 & 117.58 \\
        
        \bottomrule
    \end{tabular}
\end{table*}
\textbf{Baselines.} We do not compare against deep learning methods \cite{sun2024art,yu2023nezha,liu2025causelens}, as training and maintaining such models in our hyper-scale microservice system would be prohibitively expensive. 
% For LLM-based methods, many prior RCA systems are closely tied to proprietary production platforms or custom operational workflows, making fair reproduction in our setting difficult. 
We therefore compare \name against four representative LLM-based baselines: three widely used agent reasoning frameworks and one RCA-focused agent framework. 
(1) CoT \cite{wei2022chain} improves diagnostic accuracy by breaking the task into intermediate reasoning steps;
(2) ReAct \cite{yao2022react} combines reasoning with action, allowing the model to interact with the environment while solving the task;
(3) Reflexion \cite{shinn2023reflexion} builds on ReAct by adding self-reflection, so the agent can critique and revise earlier reasoning traces;
(4) RCA-Agent, introduced in OpenRCA \cite{xu2025openrca}, is designed for RCA and uses code-assisted data retrieval and analysis to reduce the burden of long-context diagnosis. 
All baselines are given the same diagnostic tools and the same historical knowledge base as \name, ensuring a fair comparison.

\textbf{Implementation.} The tool integration between agents and diagnostic tools is built upon the Model Context Protocol (MCP), using the open-source library\footnote{https://github.com/mark3labs/mcp-go}. For the API-level drilldown, we set the propagation threshold in Eq.~\ref{eq:score_total} to $0.8$, the maximum time lag $L=5$. The latency score weights $(w_1, w_2, w_3 )$ in Eq.~\ref{eq:score_latency} are set to $(0.2, 0.5, 0.3)$. We retain the top-$3$ suspicious services before passing them to later modules, balancing diagnostic accuracy and end-to-end latency. For the memory retrieval weights $(\alpha, \beta, \gamma)$ in Eq.~\ref{eq:rag}, we set $(0.5, 0.2, 0.3)$ for Factual Memory and $(0.4, 0.5, 0.1)$ for Experiential Memory. The 300 failure cases are used only for evaluation. All system parameters and the tiered memory are configured with routine daily alerts after removing failure-related alerts, ensuring that none of the evaluated incidents is involved in tuning.
% \textbf{Implementation.} All LLM capabilities in \name are accessed via external API services. The tool integration between agents and diagnostic tools is built upon the Model Context Protocol (MCP), using the open-source library\footnote{https://github.com/mark3labs/mcp-go}. In the API-level drilldown module, we keep only the top three suspicious services ranked by the score before passing them to later modules. This choice balances diagnostic accuracy and end-to-end latency in the production environment. 
% The 300 failure cases are used only for evaluation. All system parameters and the tiered memory are configured with routine daily alerts after removing failure-related alerts, ensuring that none of the evaluated incidents is involved in tuning.

\subsection{RQ1: Overall Performance}
\label{sec:overall performace}

Table \ref{tab:overall_exp} reports the diagnostic performance of \name and the baseline methods on real-world failures collected in Kuaishou's production environment. \name consistently outperforms all baselines in both root-cause service localization and failure-type classification, achieving AC@1 scores of 0.88 and 0.79, respectively. Compared with RCA-Agent \cite{xu2025openrca}, the strongest baseline, these results correspond to absolute gains of 31\% and 32\%, respectively. We also evaluate different seed LLMs within \name by varying the underlying foundation model. The results show that \name preserves strong performance across different LLMs relative to the baselines, further demonstrating the robustness and generalizability of the framework.

Although CoT \cite{wei2022chain} benefits from the reasoning capability of LLMs, it performs well only on relatively simple failure cases, such as cases with no more than three anomalous services. ReAct \cite{yao2022react} and Reflexion \cite{shinn2023reflexion} can progressively narrow the search space through step-by-step reasoning, but their unconstrained reasoning process requires frequent tool calls and multiple interaction rounds, sharply increasing both latency and error rate. RCA-Agent \cite{xu2025openrca}, a purpose-built RCA framework, alleviates the long-context burden through code-assisted data retrieval and outperforms the general-purpose agent baselines. Nevertheless, it still lacks structural priors to manage the massive search space in hyper-scale systems, and its effectiveness is constrained by the model's error-handling capability during code execution. We further observe that in complex cases involving more than 100 anomalous services, all agent-based baselines frequently encounter context explosion or execution failures, causing the LLM to terminate without producing an answer. In contrast, \name follows a multi-stage diagnostic paradigm that systematically reduces the search space before LLM reasoning, enabling effective handling of complex failures while maintaining high accuracy in both root cause service localization and failure type classification.
\begin{figure*}[!t]
    \centering
    \begin{subfigure}[t]{0.54\textwidth}
        \centering
        \includegraphics[width=\textwidth]{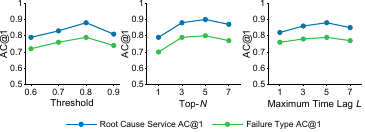}
        \caption{Impact of the threshold, Top-$N$ and the maximum time lag $L$ in the API-level drilldown module on root cause service and failure type AC@1.}
        \label{fig:hyperparameter_subfig1}
    \end{subfigure}
    \hfill
    \begin{subfigure}[t]{0.45\textwidth}
        \centering
        \includegraphics[width=\textwidth]{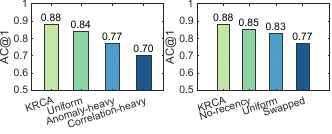}
        \caption{Impact of balancing weights in the latency score (Eq.~\ref{eq:score_latency}) and memory retrieve (Eq.~\ref{eq:rag}) on root cause service AC@1.}
        \label{fig:hyperparameter_subfig2}
    \end{subfigure}
    \vspace{-1em}
    \caption{Sensitivity analysis of key hyperparameters in \name.}
    \label{fig:hyperparameter_mainfig}
\end{figure*}
\subsection{RQ2: Ablation Study}
To validate the necessity of the key design choices in \name, we conduct an ablation study in which each module is removed or simplified while the rest of the system remains intact. In \textit{w/o API-level drilldown}, we replace API-level drilldown with service-level dependency traversal and pass all downstream services directly to the later modules. In \textit{w/o Skeleton Graph}, we remove causal graph generation and provide the agents with raw anomalous metrics without explicit structure. In \textit{w/o Multi-agent}, a single agent performs the entire reasoning process instead of multiple collaborating agents. In \textit{w/o Tiered Memory}, the memory module is disabled, so agents rely only on the current incident and cannot retrieve similar past failures or their associated reasoning insights. In \textit{w/ Naive RAG}, we replace the structured tiered retrieval scheme with flat similarity-based retrieval over a single memory pool that merges all memory types. We also set all weights in Eq.~\ref{eq:rag} to $1/3$, so retrieval no longer distinguishes among different entry types. The results are reported in Table~\ref{tab:ablation}.

Without API-level drill-down, the system cannot filter services finely along invocation paths, so many irrelevant services remain in the candidate set. This mainly hurts root cause service localization, while failure type classification is less affected because the failure type is often still visible in upstream metrics. Without the skeleton graph, agents must reason over raw anomalous metrics without prior structure, which increases hallucination and reduces accuracy for both AC@1 root cause localization and failure type classification. Without multi-agent collaboration, all reasoning is handled by a single agent, which is less effective at verifying causality across heterogeneous metric categories. This leads to consistent degradation on both tasks. Disabling the memory module also reduces localization and classification performance, because the LLM can no longer draw on relevant historical cases and must reason from scratch. Replacing the proposed composite RAG with Naive RAG similarly causes clear performance drops in both tasks, because poorly matched retrieved cases mislead the LLM and divert its reasoning away from the true root cause.

\subsection{RQ3: Hyperparameter Study}
\label{sec:hyperparameter}
\begin{table}[!t]
\centering
\caption{Ablation study of \name.}
\label{tab:ablation}
\setlength{\tabcolsep}{3.5pt}
\small
\resizebox{\columnwidth}{!}{
\begin{tabular}{lcccc}
\toprule
\multirow{2.5}{*}{\textbf{Methods}} 
& \multicolumn{2}{c}{\textbf{Root Cause Service}} 
& \multicolumn{2}{c}{\textbf{Failure Type}} \\
\cmidrule(lr){2-3} \cmidrule(lr){4-5}
& \textbf{AC@1} & \textbf{AC@3} & \textbf{AC@1} & \textbf{AC@3} \\
\midrule
\textit{\name} 
& \textbf{0.88} & \textbf{0.96} & \textbf{0.79} & \textbf{0.98} \\
\textit{w/o API-level drilldown} 
& 0.64 & 0.82 & 0.76 & 0.95 \\
\textit{w/o Skeleton Graph} 
& 0.72 & 0.91 & 0.61 & 0.92 \\
\textit{w/o Multi-agent} 
& 0.75 & 0.88 & 0.67 & 0.89 \\
\textit{w/o Tiered Memory} 
& 0.81 & 0.93 & 0.74 & 0.95 \\
\textit{w/ Naive RAG} 
& 0.78 & 0.91 & 0.71 & 0.93 \\
\bottomrule
\end{tabular}
}
\vspace{-1em}
\end{table}

To assess the stability of \name, we perform a sensitivity analysis of its key hyperparameters. The analysis covers three parameters in the API-level drilldown module (Fig.~\ref{fig:hyperparameter_mainfig}(a)) and two sets of balancing weights (Fig.~\ref{fig:hyperparameter_mainfig}(b)). For the drilldown module, the threshold in Eq.~\ref{eq:score_total} gives AC@1 values ranging from 0.79 to 0.88, with the best result at 0.8. Top-$N$ produces AC@1 between 0.79 and 0.90, and all settings with $N \ge 3$ remain above 0.87. The maximum time lag $L$ in Eq.~\ref{eq:score_total} is the most stable parameter, with AC@1 staying within 0.82 to 0.88 for $L \in {1, 3, 5, 7}$. Although $N=5$ achieves the highest AC@1 of 0.90, it also incurs much higher latency than $N=3$ (191.2s vs.\ 146.6s). We therefore use Top-3 in production to balance diagnostic accuracy and real-time efficiency. For the latency score weights in Eq.~\ref{eq:score_latency}, we compare \name $(0.2, 0.5, 0.3)$, Uniform $(1/3, 1/3, 1/3)$, Anomaly-heavy $(0.6, 0.2, 0.2)$, and Correlation-heavy $(0.2, 0.2, 0.6)$. The corresponding AC@1 values are 0.88, 0.84, 0.77, and 0.70, which suggests that the system remains stable unless correlation is given excessive weight. For the memory retrieval weights in Eq.~\ref{eq:rag}, we compare \name (Factual: $(0.5, 0.2, 0.3)$, Experiential: $(0.4, 0.5, 0.1)$), Uniform (both $(1/3, 1/3, 1/3)$), Swapped (Factual: $(0.4, 0.5, 0.1)$, Experiential: $(0.5, 0.2, 0.3)$), and No-recency (Factual: $(0.7, 0.3, 0.0)$, Experiential: $(0.4, 0.6, 0.0)$). Their AC@1 values are 0.88, 0.83, 0.77, and 0.85, respectively. Excluding the structurally mismatched Swapped setting, performance stays within a narrow range of 0.83--0.88. Even under the least favorable configuration (\eg Correlation-heavy, AC@1 $= 0.70$), \name still outperforms all baselines reported in Section~\ref{sec:overall performace}.

\subsection{RQ4: Usability for SREs}

To evaluate the practical utility of the failure reports generated by \name in real-world failure scenarios, we invited five SREs, each with at least five years of hands-on operational experience, to assess the reports. We randomly sampled 30 failure cases and adopted the same scoring criteria as \cite{zhaollms}, namely a 5-point Likert scale where 1 indicates the lowest quality and 5 indicates the highest. All SREs rated the reports independently to ensure the objectivity of the evaluation.
As shown in Table~\ref{tab:utility}, the average readability score reached 4.21, with a HIP (the percentage of samples rated 4 or above \cite{liu2024logprompt}) of 83.3\%, indicating that the majority of reports are clearly structured and easy to comprehend. The average usefulness score reached 4.10, with a HIP of 79.3\%, suggesting that most reports effectively assisted SREs in their troubleshooting process. These results, grounded in firsthand feedback from frontline SREs, confirm that \name provides tangible support during failure diagnosis in production environments.

\begin{table}[!t]
\centering
\caption{Practical utility of \name rated by SREs.}
\label{tab:utility}
\begin{tabular}{ccccc}
\toprule
\multirow{3}{*}{\textbf{SREs}} & \multicolumn{2}{c}{\textbf{Readability}} & \multicolumn{2}{c}{\textbf{Usefulness}} \\ 
\cmidrule(lr){2-3} \cmidrule(lr){4-5}
 & Mean & HIP & Mean & HIP \\ 
 
\midrule
SRE \#1 & 4.30 & 86.67\% & 4.20 & 83.33\% \\
SRE \#2 & 4.27 & 86.67\% & 4.17 & 83.33\% \\
SRE \#3 & 4.20 & 83.33\% & 4.10 & 80.00\% \\ 
SRE \#4 & 4.17 & 80.00\% & 4.03 & 76.67\% \\ 
SRE \#5 & 4.11 & 80.00\% & 4.00 & 73.33\% \\ 
\midrule
\textbf{Avg.} & \textbf{4.21} & \textbf{83.33\%} & \textbf{4.10} & \textbf{79.33\%} \\ 
\bottomrule
\end{tabular}
\vspace{-1em}
\end{table}

%% file: Chapter/Discussion.tex
\subsection{Deployment}
\begin{figure}[t]
    \centering
    \includegraphics[width=\linewidth]{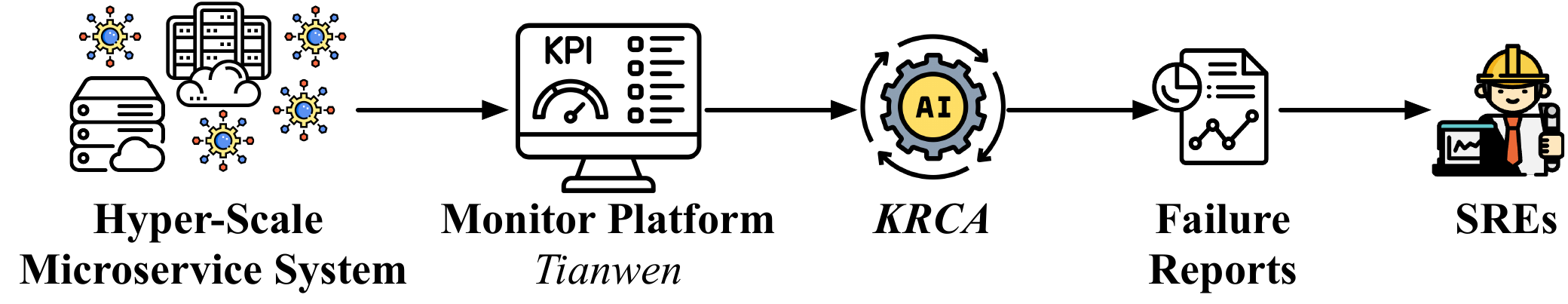}
    \caption{Deployment architecture of \name.}
    \label{fig:deployment}
    \vspace{-10pt}
\end{figure}

\name has continuously deployed in Kuaishou's internal monitoring platform, \textit{Tianwen}, from October 2025 to March 2026. As shown in Fig.~\ref{fig:deployment}, it currently provides RCA support for more than 2{,}000 alerts per day across six major business lines, including Short Video, E-commerce, and Algorithms. During this period, we collected internal statistics on 483 system-related emergency incidents\footnote{System-related emergency incidents refer to incidents caused by anomalies in system operation, such as abnormal traffic, CPU/GPU saturation, database performance degradation, and service dependency failures. An emergency incident does not necessarily develop into a failure if it is mitigated promptly and causes only limited impact.}. For each incident, the ground-truth root cause service and failure type were established through postmortem analysis by the SREs responsible for the affected services. According to these records, \name correctly identified both the root cause service and the failure type in 82\% of incidents, substantially reducing diagnosis effort in practice. We also analyzed the 18\% of incidents for which \name failed to produce a correct diagnosis. These failures mainly fall into two categories. Firstly, some key downstream services lacked critical observability signals, such as availability-related metrics, which caused the drilldown process to terminate too early. Secondly, multiple cascading anomalies occurred simultaneously, causing several services to fail at once and making it difficult to separate the true propagation path from correlated noise. 

Before the deployment of \name, incident diagnosis at Kuaishou mainly relied on manual, layer-by-layer inspection. Across all failure cases collected between 2024 and October 2025, the average time required to localize the root cause was about 52 minutes. After deployment, this average fell to 11.8 minutes for failures observed from October 2025 to March 2026, a 77.3\% reduction. In our annual user survey, \name received a Net Promoter Score (NPS) of 74\%, indicating strong satisfaction among SREs. 
% In production, we use Claude-opus-4.6 as the Main Agent and GPT 5.1 as the Sub Agents. On average, \name consumes 2.3801 billion tokens per month.  On average, \name consumes 2.3801 billion tokens per month.

\subsection{Case Study}

To show how \name performs in practice, we present a real cascading failure from Kuaishou's production environment. The failure began with a coredump in a far-downstream storage service, storage\_A. This failure triggered connection timeouts that propagated across three layers of dependencies and eventually appeared as a severe availability drop in the upstream recommendation service reco\_B. Each layer contained dozens of candidate services, and several intermediate services showed different levels of availability degradation, making it difficult to separate the true propagation path from background noise. After the alert was triggered, \name automatically followed the anomaly along API invocation paths, quickly ruled out irrelevant branches, and identified storage\_A as the primary suspect. It then organized the anomalous metrics into a structured causal graph and invoked domain-specific Sub Agents to collect supporting evidence. By combining the collected evidence, the agents found a sudden process termination in host-level logs and matched the case to similar historical incidents stored in the memory system, which confirmed a memory access violation as the root cause. The resulting diagnostic report was immediately sent to the responsible SREs, allowing rapid mitigation.

For this incident, before \name was deployed, diagnosing a similar cascading failure typically required about 30 minutes. With \name, the end-to-end diagnosis time for this case was reduced to about 5 minutes. Since the availability degradation of reco\_B is estimated to cause a Gross Merchandise Volume (GMV) loss of \$5{,}200 per minute, this improvement corresponds to an estimated reduction of more than \$130{,}000 in potential business losses for this single incident.

\subsection{Lessons Learned}

\textbf{Context engineering.} We have invested significant effort in optimizing the LLM context within \name—including prompts and tool definitions—and have gained several key insights. First, managing context length is vital for balancing efficiency and accuracy. We have observed a noticeable decline in reasoning precision as prompt length increases, coupled with a sharp rise in inference latency. To mitigate this, we decompose complex logic into multiple sub-agents (Section~\ref{sec: which}) to keep individual prompts concise, while utilizing parallel execution to reduce end-to-end delay.  Second, the robustness of tool invocation is critical for maintaining a stable reasoning trajectory. We found that erroneous or malformed tool returns can severely derail the multi-agent system, leading to outcomes that diverge significantly from expectations. Since different agents often interact with the same tool (\eg metric queries) using heterogeneous parameter schemas, providing generic descriptions is often insufficient. To ensure precise calls, we provide specialized tool versions and role-specific descriptions tailored to each agent’s specific context. We have further enhanced this with techniques such as error-handling loops to safeguard the integrity of the diagnostic process.

\noindent \textbf{Explainability.} \name provides three key features to enhance result interpretability and assist SREs in troubleshooting: (1) link topology visualization, where a graph of suspicious downstream services is delivered immediately after API-level drilldown (Section~\ref{sec: where}) to provide a holistic view of the failure (latency $<$ 5 s); (2) anomaly detection results (Section~\ref{sec: what}), which are performed concurrently across all suspicious services and surfaced in the \textit{Tianwen} system to minimize the manual effort of inspecting metrics (latency $<$ 10 s); (3) comprehensive failure reports, generated by the multi-agent system (Section~\ref{sec: which}) to offer a deep-dive failure analysis (latency $<$ 3 mins). This design philosophy is rooted in progressive context disclosure. We have found that by providing diagnostic information in stages, SREs can obtain actionable insights at different time intervals, which significantly increases the adoption rate and operational trust in \name during real-world incidents.

%% file: Chapter/Related_work.tex
\textbf{Causal discovery.} Causal discovery seeks to infer causal relations from observational data. Existing approaches generally fall into three categories: (1) statistical methods based on conditional independence tests (\eg PC algorithm and Granger causality) \cite{chen2014causeinfer, ma2020automap, meng2020localizing, ikram2022root, lin2024root, cheng2024cuts+}; (2) deep learning approaches, especially GNN-based methods, that learn causal structures from evolving topologies \cite{wang2023incremental, yu2023nezha}; and (3) recent LLM-based methods that extract causal intent from text \cite{du2025causal, susanti2025paths, yang2025causalmob}.
However, these approaches exhibit fundamental limitations in hyper-scale microservice systems. Statistical methods are vulnerable to high-dimensional and coarse-grained monitoring data, and they often produce spurious relations when unrelated metrics surge simultaneously. Deep learning models struggle to remain effective as system scale increases, because both noise and dimensionality grow substantially. At the same time, the direct use of LLMs on massive time-series data is neither practical nor scalable.
In contrast, \name addresses these limitations by instantiating a high-recall skeleton graph as a structural prior instead of inferring causality directly from raw time-series data, and then leveraging a multi-agent framework to verify fine-grained causal relations with rich metric semantics.

\textbf{LLM-based RCA.} The rapid progress of LLMs has inspired many automated RCA frameworks \cite{liu2025r, shi2025flowxpert, sun2025llm, zhaollms}, including tool-augmented autonomous agents (\eg RCAgent \cite{wang2024rcagent}, Flow-of-Action \cite{pei2025flow}), in-context learning systems (\eg RCACopilot \cite{chen2024automatic}), and domain-specific diagnostic bots (\eg D-Bot \cite{zhou2023d}, OpenRCA \cite{xu2025openrca}, TrioXpert \cite{sun2025trioxpert}).
Although promising, these methods face severe scalability challenges in hyper-scale systems. They typically place LLMs directly in the RCA loop and expose them to a large and noisy search space from the beginning. As a result, the LLMs must depend on extensive multi-round reasoning and repeated tool use to narrow the candidate set, which significantly increases end-to-end latency and the risk of hallucination.
This scalability bottleneck highlights the need for a more structured approach. \name addresses this issue by applying an efficient API-level drilldown strategy to sharply reduce the search space before introducing LLMs, thereby ensuring both real-time diagnostic efficiency and high accuracy in production environments.

%% file: Chapter/Conclusion.tex
In this paper, we present \name, an end-to-end root cause analysis system designed for hyper-scale microservice systems. To address the central challenges posed by extreme dynamism and massive scale, \name adopts a progressive multi-stage diagnostic paradigm. It first applies an API-level drilldown strategy to effectively reduce the search space and identify suspicious services. It then constructs a skeleton-based causal graph from anomalous metrics, which provides a high-recall structural prior for subsequent analysis. Finally, \name employs a memory-augmented multi-agent framework to collaboratively verify causal relationships and generate comprehensive failure reports. Extensive experiments on a dataset of 300 real-world failures collected from the production environment of Kuaishou show that \name significantly outperforms state-of-the-art baselines in both root cause service localization and failure type classification. In addition, \name has been deployed in a real-world production environment for more than six months, where it processes over 2,000 alerts per day and substantially reduces MTTR.